\documentclass[fleqn]{article}

\usepackage{graphicx}
\usepackage{epsfig}

\begin{document}

\newcommand{\beq}{\begin{equation}}
\newcommand{\eeq}{\end{equation}}
\newcommand{\rang}{\rangle}
\newcommand{\lang}{\langle}
\newcommand{\ovr}{\overline}
\newcommand{\mna}{\mu \nu \alpha}

\title{{\bf Communication and measurement \\
with squeezed states}}

\author{Horace P. Yuen\\
Department of Electrical and Computer Engineering,\\
Department of Physics and Astronomy,\\
Northwestern University, Evanston, IL 60208}
\maketitle

\begin{abstract}
The principles are elaborated which underlie the applications of
general nonclassical states to communication and measurement systems.
Relevant classical communication concepts are reviewed.  Communication
and measurement processes are compared.  The possible advantages of
nonclassical states in classical information transfer are assessed.
The significance of novel quantum amplifiers and duplicators in
communication is emphasized.  A general approach is developed for
determining the ultimate accuracy limit in quantum measurement
systems.  It is found that bandwidth or mode number is a most
important parameter and ultrahigh precision measurement is
possible in systems with a fixed energy but many modes.  The problem
of the standard quantum limit in monitoring the position of a free
mass is also addressed.
\end{abstract}

\vfill

\noindent{To appear in the book {\em Quantum squeezing}, P.D. Drummond and
Z. Ficek, Springer-Verlag, to be published.}

\newpage
\tableofcontents
\newpage

\section{Introduction}\label{ysec.1}

In this chapter I will discuss the advantages, in principle, of using
nonclassical states [1] in communication and
measurement situations involving classical information transfer.  Most
of the discussions will be concerned with the quantum states of light,
in particular, the quadrature squeezed states and number states.
Thus, optical terminology will be freely employed even though the
principles are generally applicable to fermions also, and
gravitational wave detection by a free mass will also be treated.  I
shall focus on the general theoretical concepts and principles
underlying such applications of nonclassical states without extensive
mathematical derivations, and also no review of the physics involving
these states which are covered elsewhere in this book.  I shall
mostly avoid precise mathematical definitions and formulations,
although the treatment is as precise as most standard treatments in
theoretical physics or engineering science.

It is, of course, the defining characteristic of a quadrature squeezed
state that the quantum fluctuation in one of its quadrature is reduced
below that of a coherent state.  Let $|\alpha \rang$ be a coherent
state (CS) of an optical field mode with photon annihilation operator
$\hat{a}= (\hat{x} + i\hat{y})/2$ so that
\beq
\left(\Delta x \right)^{2} = \left(\Delta y \right)^{2} = 1.
\label{y1}
\eeq
In a two-photon coherent state (TCS) [2] $|\mna
\rang$, which are the pure quadrature squeezed states, one obtains
with a proper choice of quadrature $\hat{x}^{\theta}$
\beq
\left(\Delta \hat{x}^{\theta} \right)^{2} = (|\mu| - |\nu |)^2, \qquad
\left(\Delta \hat{x}^{\theta +\pi/2} \right)^{2} = (|\mu| + |\nu |)^2.
\label{y2}
\eeq

\vspace{2ex}
Since $|\mu |^2 - |\nu |^2 = 1$, $|\mna \rang$ is a minimum
uncertainty state on $\hat{x}^{\theta}$, $\hat{x}^{\theta +\pi/2}$,
\beq
\left(\Delta x^{\theta}\right)^{2}\left(\Delta
x^{\theta+\pi/2}\right)^{2} =1. \label{y3}
\eeq
\vspace{2ex}
For simplicity, let $\mu, \nu, \alpha$ be real and $|\mu - \nu
| < 1$, thus
\beq
\lang \alpha |\hat{x}|\alpha \rang = 2\alpha = \lang \mna |\hat{x}|\mna \rang
\label{y4}
\eeq
\beq
\lang \alpha |\left(\Delta \hat{x}\right)^{2}|\alpha \rang >
\lang \mna |\left(\Delta \hat{x}\right)^{2} |\mna \rang . \label{y5}
\eeq
\vspace{2ex}
This is often taken to mean that in the proper quandrature, a TCS is
less noisy than a CS and so is better for communication and
measurement.  However, (\ref{y4})-(\ref{y5}) is {\em not} a proper
justification of such an assertion.

First of all, the states $|\mu \nu \alpha \rang$ and $|\alpha \rang$,
which is $|\mna \rang$ with $\nu = 0$, have different
energy,
\beq
\lang \mna | a^{\dagger}a|\mna \rang = |\alpha |^2 + |\nu |^2. \label{y6}
\eeq
\vspace{2ex}
It is not a priori clear that if a portion of the energy associated
with the mean field $\alpha$ is moved to increase
$\left(\Delta y\right)^{2}$ so that $\left(\Delta x\right)^{2}$ is less,
the overall effect is beneficial.  Assuming that a signal-to-noise ratio
(SNR) criterion is appropriate for the present illustration,
\beq
\mbox{SNR} \equiv \frac{\left\langle x\right\rangle^{2}}
{\left(\Delta x\right)^{2}}
\label{y7}
\eeq
\vspace{2ex}
it was shown [3] that TCS indeed maximizes (\ref{y7}) under the
constraint of a fixed energy for an arbitrary state, $\lang
a^{\dagger}a\rang \leq S$, with the result
\beq
\mbox{SNR}_{|\mna \rang} = 4S(S+1) \label{y8}
\eeq
\vspace{2ex}
for $\nu = S/\sqrt{2S+1}$ as compared to SNR$_{|\alpha\rang}$ = $4S$.
Secondly, in communication with $|\alpha\rang$, both quadratures can be
used to carry information and thus may yield a higher capacity than the
use of $|\mna \rang$ with only one quadrature, which is equivalent to
using half of the available bandwidth.  It turns out that for the
unrestricted capacity [4], and much more so for the
binary signaling capacity [5], the use of $|\mna
\rang$ does lead to improvement over $|\alpha \rang$.  The relevant
communication concepts and further details are to be discussed in the
sequel.  The point here is that the advantage of $|\mna \rang$ over
$|\alpha \rang$ is not as obvious or intuitive as it may first appear.
Similarly, while number states $|n\rang$ and direct detection produce
a noiseless system, it is discrete as compared to the in-principle
continuum of states $|\alpha\rang$.  Again, it is not a priori clear
that $|n\rang$ would lead to a higher capacity.

The real point involving nonclassical states, I believe, is the
following.  Historically or typically in physics, one analyzes a given
physical phenomenon and sees if it can be useful in application,
whereas in engineering one often synthesizes to produce something to
perform a certain function efficiently.  (This opposition between
analysis and synthesis is, of course, neither absolute nor pervasive
in physics versus engineering.)  For a long time after the laser was
invented, the ideal laser state was supposed to be a coherent state, a
quantum source one has to live with.  Thus, all practical light
sources were supposedly characterized by classical states, i.e., pure
coherent states or their random superposition.  However, states which
are not classical, the {\em nonclassical states}, are clearly possible
to have, at least in principle.  In a synthesis or optimization
approach, one would want to find out whether such states could lead to
a better system for the application under consideration. Thus, the
following {\em questions} suggest themselves in any given problem
situation: What are the appropriate performance criteria and resource
constraints?  What are the best states or state-measurement
combination one should use according to the criteria and the
constraints?  How much better are they compared to the conventional or
standard system?  The above discussion surrounding (\ref{y7}) and (\ref{y8})
furnishes
an example of answers to such questions. Typically, the answer would
involve quantities that are only specified mathematically, such as a
TCS. If it seems worthwhile to develop such new systems, further
questions on concrete physical realizations would have to be
addressed.  In these days of ``quantum information'', such questions are 
even more pervasive and important.

In the following section, I will review some basic concepts in classical
communication, distinguish communication from detection, and discuss how
physical measurement fits into both. In Section~\ref{ysec.3} the
issues of quantum communication for classical information transfer
will be explained. [Note that ``quantum
information'' is entirely outside the scope of this chapter.]  I will
discuss the information capacity of nonclassical states, and the
apparently only possible useful application of nonclassical states in
fiber optic communication, to date --- the use of nonclassical amplifiers and
duplicators. In Section~\ref{ysec.4}, I will discuss the use of nonclassical
states in physical measurement problems, and the communication
theoretic limit on the accuracy of measurements. In Section~\ref{ysec.5}
the validity of the standard quantum limit for monitoring free-mass
positions is addressed.  Throughout I will try to explain the
intuitive relevance of the various basic communication parameters, to
highlight the main ideas with careful formulation but minimum details, 
and to dispel a few
common misconceptions. Some results are also presented here for the
first time.

\section{Classical communication and measurement}\label{ysec.2}

\subsection{Classical Information Transmission}\label{ysec.2.1}

\begin{figure}
\includegraphics[width=\textwidth]{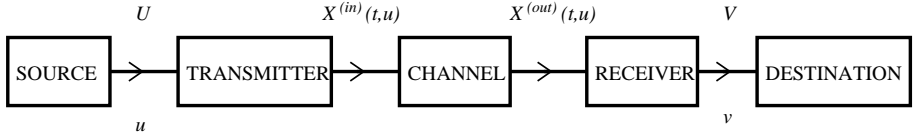}
%%\centering\psfig{file=./yue1.eps,width=.9\textwidth}  
\caption{Schematic representation of a classical communication 
system: for $U$ and $V$ capitals denote random quantities, 
lower case their samples but no such distinction is made for $X^{(in)}$
and $X^{(out)}$.}
\label{yufig1}
\end{figure}

For our purpose, a classical communication system can be schematically
represented by Fig.~\ref{yufig1}. A source generates a classical
quantity $u$,
which is a member of an alphabet set U, $u \in $U, which may be
discrete or continuous.  Since $u$ is generated probabilistically
according to some distribution, the corresponding random variable is
denoted by $U$.  The transmitter modulates $u$ onto a signal $X^{(in)}(t,u)$,
which is a time-varying classical function.  The channel, which
usually represents all the disturbance in the system from source to
destination, yields an output $X^{(out)}(t,u)$ statistically related to the
input $X^{(in)}(t,u)$.  The receiver processes $X^{(out)}(t,u)$ to produce an
estimate $v \in$ U of $u$ to satisfy the performance criteria.

If U is a finite set $\{ 1, \cdots, M \}$, the criterion of error
probability is often employed.  If U is continuous, the mean-square
error between $U$ and $V$ is often taken as the criterion.  In both
cases the system is designed, subject to whatever constraints under
consideration, to minimize the error or to produce a sufficiently
small error.  In a {\em communication} situation, one has joint design
over the transmitter and receiver whereas in a {\em detection}
situation, one is concerned only with the receiver design.  Thus, in
communications one may pick $X^{(in)}(t,u)$ to influence $X^{(out)}(t,u)$,
and in detection one is faced with a given statistical description of
$X^{(out)}(t,u)$.  Clearly, communication is broader than detection.  In the
communication case it is important to deal explicitly with the
time-sequential nature of the source output, with $u$ regarded as a
sequence $u_1, u_2, \cdots, u_i, \cdots$ with corresponding
$X^{(in)}_i(t,u_i)$ and $X^{(out)}_i(t,u_i)$.

The system constraints in both cases are similar. The physical
transmission medium (and often together with the unavoidable
disturbance in the receiver structure) specifies the {\em channel
representation}, the statistical relation between $X^{(in)}(t,u)$ and
$X^{(out)}(t,u)$. Constraints on the channel typically include all the
physical limitations on the transmitter, the medium, and the receiver.
They usually include a power or energy limitation on $X^{(in)}(t,u)$, a total
time $T$ and a total bandwidth $W$ available for transmission and
reception.  In addition to small error, the system objectives include
moderate implementation complexity, which is not always easy to
quantify, and also large data rate in the case of communication.

The concept of data rate or information rate is fundamental in
communication. It is usually measured in bits per second, or bits per
use which is immediately converted to bits per second when multiplied
by uses per second. For a data source generating one of $M$
equiprobable messages per $T$ seconds, the data rate $R$ is defined to
be
\beq
R = (\log _2 M)/T . \label{y9}
\eeq
This definition explicitly indicates that it is the {\em number of
message possibilities} that characterizes the rate of a source.  It
immediately shows why one can have more than one bit per photon.
Indeed, one can have an infinite number of bits per photon if that
photon can fall into, say, one of an infinite number of different time
slots.  For a general statistical source, the Shannon entropy $H$ for
the source is used, in bits per use of the source or bits per
source symbol.  A full description of communication, information, and
detection theory can be found in [6]-[9]. In the present treatment,
only some significant relevant points would be highlighted.

The concept of data rate (\ref{y9}) already forces upon us a {\em fundamental
discrete view of nature} in any realistic physical process.  If one
can assess a true continuum, or indeed a true discrete infinity (in
communications the word ``discrete'' often means discrete and finite),
one would be able to get infinite data rate, e.g., when one can
distinguish the real numbers between 0 and 1 with infinite precision.
In reality, a continuum can support only a finite number of bits
either from unavoidable disturbance or from the laws of quantum
physics.  A discussion of certain points relating to this
finite/infinite dichotomy can be found in [10].
Here I would like to emphasize that communication is inherently a
finite (discrete, digital) process.  Any continuous quantity would
finally appear in some discrete fashion in actual utilization.

Not surprisingly, the desirable goals of large data rate and small
error probability are in conflict with each other.  It is easy to see
from the law of large number that if one slows down the data rate, say
by repeatedly sending the same message, one can decrease the error
probability, indeed to zero asymptotically but with the data rate also
going to zero.  What is not obvious, but given by Shannon's channel
coding theorem, is that for a fixed channel representation, there is a
nonzero rate called channel capacity below which one can
transmit with arbitrarily small error probability by using
increasingly long codes.  A (channel) code is a signaling scheme in
which all the signaling symbols in a sequence over many uses are
processed simultaneously, which clearly makes the implementation more
complex.  However, long sequences have the statistical regularity
given by the probabilistic description similar to the law of large
number, which, e.g., implies that in a long sequence of fair coin
tosses there is roughly one half heads, compared to nothing that can
be said which is applicable to a single or a few tosses. It is this
{\em statistical regularity in long sequences} that leads to the
possibility of vanishingly small error probability with a nonzero rate
as given by Shannon's theorem.

The maximum such nonzero rate under whatever constraints and
specifications on a channel is called the {\em capacity} of that
constrained or specific channel, and is equal to the mutual
information between the channel input and output.  Referring to
Fig.~\ref{yufig1}, we will later discuss the time-varying signal
aspect but for
the moment consider just channel input $X^{(in)}$ and output $X^{(out)}$
from alphabets X and Y with the channel specified
statistically by the conditional probability $p(X^{(out)}|X^{(in)})$,
$X^{(out)} \in$ Y, $X^{(in)} \in$ X, interpreted as a
probability density or
probability mass according to whether the alphabet is continuous or
discrete.  With an input probability $p(X^{(in)})$, the joint probability
$p(X^{(out)},X^{(in)}) = p(X^{(out)}|X^{(in)})p(X^{(in)})$ completely
specifies the channel action and
the mutual information $I (X; Y)$ is defined by, in the continuous case
\beq
I(X; Y) \equiv \int p(X^{(in)},X^{(out)}) \log
\frac{p(X^{(in)}|X^{(out)})}{p(X^{(in)})} dX^{(in)}dX^{(out)} ,
\label{y10}
\eeq
and similarly, in the discrete case,
\beq
I(X; Y) \equiv \sum_{X^{(in)},X^{(out)}} p(X^{(in)},X^{(out)})
\log \frac{p(X^{(in)}|X^{(out)})}{p(X^{(in)})} .\label{y11}
\eeq
The Shannon entropy $H$(U) of a single random variable U can be
defined as average self information, or
\begin{eqnarray}
H(U) & \equiv & -\int p(u) \log p(u)du ,\label{y12} \\
H(U) & \equiv &  -\sum_{u} p(u) \log p(u) \label{y13}
\end{eqnarray}
in the continuous and discrete case.  Note that (\ref{y13}) is always
nonnegative while (\ref{y12}) can be negative.  Shannon's source-channel
coding theorem and its converse [7,9,12] state that
successive independent samples of a discrete $U$ can be transmitted
over a memoryless channel $p(X^{(out)}|X^{(in)})$ with arbitrarily small
(but not exactly zero) error probability between $U$ and $V$
(see Fig.~\ref{yufig1})
if $H(U) < I(X;Y)$, and the block error $P_e \rightarrow 1$ for $H(U)
> I(X; Y)$.  The case $H(U) = I(X; Y)$ forms a {\em boundary} with
$P_e$ bounded away from zero in general unless the channel is
noiseless. It is important to observe the conceptual distinction
between a source output $U$ and a channel input $X$, even though they
may happen to be the same physical quantity.  Note also that a
continuous alphabet channel in reality still has a {\em finite}
capacity and so can reliably transmit only a discrete quantity.  If
$U$ is a continuous random variable, some performance criterion such
as mean-square error would need to be adopted which cannot be made
vanishingly small.  The extent to which it can be minimized is dealt
with in rate-distortion theory [7-13] discussed in
Section~\ref{ysec.4}. It is important to note that for a noisy channel,
the use of long codes to obtain a reliable system with high rate significantly
increases the system complexity, especially in the decoding operation.

The name ``capacity'' is usually applied to the $I(X; Y)$ maximized with
respect to $p(X^{(in)})$ under whatever constraints, but it is also used to
refer to whatever maximum $I(X; Y)$ obtained by different restrictions
on the utilization of a given channel, e.g., under discretization
(usually called quantization in the communication and signal
processing literature) of the input and output of a continuous
channel. The point is that with various special restrictions including
a fixed $p(X^{(in)})$, a given channel would give rise to many other
channels, each with its own ``capacity.''  Even more proliferation
occurs in the quantum case.  It is essential to understand the exact
conditions under which a so-called ``capacity'' is obtained, for it is
often not a really meaningful capacity in the sense of {\em ultimate}
capability limit on the transmission medium or system.

\subsection{Signal, noise and dimensionality}\label{ysec.2.2}

I will now try to describe qualitatively the effect of noise on data
rate, finally leading to the famous Shannon capacity formula for an
additive white Gaussian noise (AWGN) channel which is directly
applicable to squeezed states.  Let $P$ and $N$ be the total average
signal and noise power of an AWGN channel represented by
\beq
X^{(out)}(t) = X^{(in)}(t) + n(t), \label{y14}
\eeq
where $n(t)$ is the white noise.  Let $W$ be the available bandwidth,
i.e., the duration in frequency occupied by the signals $X^{(in)}(t)$.
Then the optimizing input signals for capacity is a white Gaussian process
with resulting
\beq
C = W \log (1 + \frac{P}{N}). \label{y15}
\eeq
In terms of the noise spectral density $N_0$, one has the famous formula
\beq
C = W \log (1 + \frac{P}{N_0W}). \label{y16}
\eeq
Equation (\ref{y15}) can be derived from the mutual information expre\-ssion
(\ref{y10}), as given by Shannon [11] and later more rigorously in~[7].
The intuitive reason why (\ref{y15}) takes the form it does,
according to such a derivation, would then have to be traced through
the reason why (\ref{y10}) or (\ref{y11}) provides a {\em general} capacity for
information transmission.  In the discrete case (\ref{y11}), this can be
gleaned from Shannon's original proof in~[11], and the continuous
case may be viewed as a discrete limit as developed in~[7].
However, a direct approach can be given for a Gaussian channel, also
provided by Shannon~[13], which explains the nature of various
relevant quantities quite succinctly.

Consider the transmission and reception of a single continuous real
variable in noise
\beq
X^{(out)} = X^{(in)} + n . \label{y17}
\eeq
If $X^{(in)}$ is restricted to an interval of length $L$, an infinite number
of bits per channel use is obtained in the absence of noise, $n=0$, for
any $L > 0$.  If the noise $n$ always has value in the interval
$[-\Delta /2, \Delta /2]$, the number of bits per use is
reduced to the finite
\beq
(L + \Delta )/\Delta \label{y18}
\eeq
including edge effects. If $X$ and $n$ are independent continuous
random variables with variances $P$ and $N$, or standard deviations
$\sqrt{P}$ and $\sqrt{N}$, a crude estimate patterning after (\ref{y18})
would suggest that the number of amplitudes that can be well
distinguished, or equivalently the number of bits per use, is
\beq
\sim k \sqrt{(P+N)/N} ,\label{y19}
\eeq
where $k$ is a small constant in the neighborhood of unity depending
on how ``well distinguished'' is to be interpreted.  We may recall
earlier in this section it was mentioned that in a long sequence of
independent trials, statistical regularity appears and provides
deterministic features to the sequence.  This kind of effect would
indeed turn the approximate relation (\ref{y19}) into an exact one similar
to (\ref{y18}).  In the case of time-varying signals, this comes about in the
long signal duration $T$ limit as follows.

First of all, the collection of time functions of ``approximate time
duration'' $T$ and ``approximate bandwidth'' $W$ span a linear space
of dimension
\beq
D \sim 2TW \label{y20}
\eeq
according to the Dimensionality Theorem [6], an improved version
of the sampling theorem [13].  The word ``approximate'' above is
necessary because no time function with a Fourier transform can be
both strictly time-limited and strictly band-limited, but the exact definitions
of ``approximate'' do not alter the final result (\ref{y20}) [14].  The
Dimensionality Theorem (\ref{y20}), as I discussed elsewhere [15], has momentous
consequences in the description of nature.  Here, it cuts
down, even in the absence of noise, an otherwise infinite dimensional
space to a finite dimension in a realistic system where both $T$ and
$W$ have to be finite.  Thus, a signal or time function can be viewed
geometrically as a point in a finite dimensional Hilbert space.  (The
linear space is readily given an inner product via $\int_T X^{(in)}_1(t)
X^{(in)}_2(t)dt$ .)

\begin{figure}
\includegraphics[width= \textwidth]{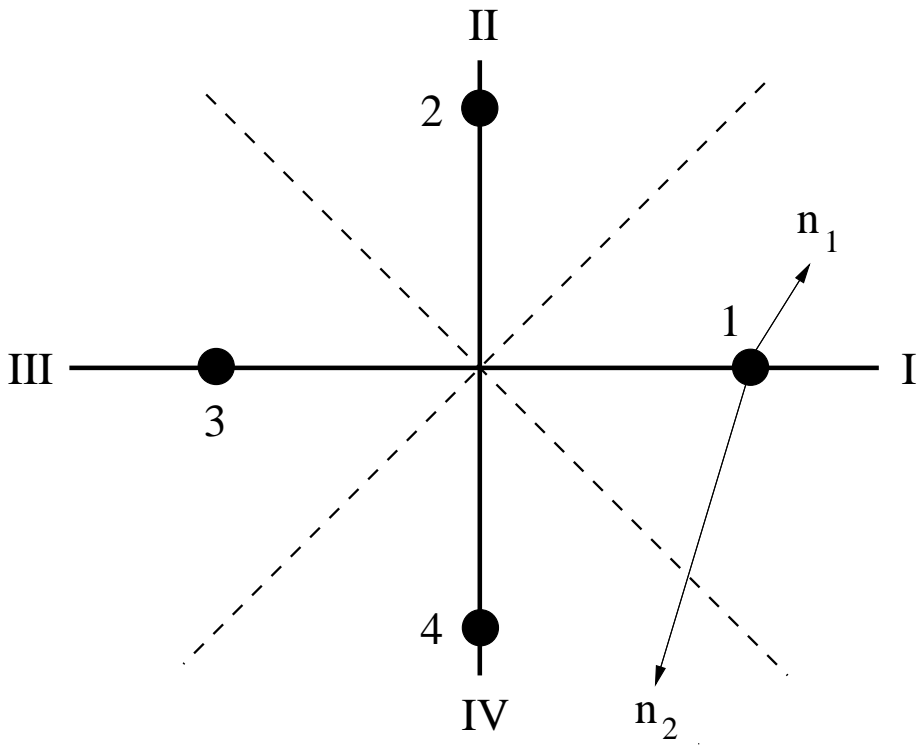}

\caption{Geometric representation of the receiver: four possible
signals 1 to 4 in a 2-dimensional space with corresponding minimum
distance decision regions I to IV formed by the dotted lines.  The
additive white Gaussian noise vector $n_1$ added to signal 1 would
be decoded correctly, while the noise vector $n_2$ pushes signal 1
to the decision region IV for signal 4, and would be decoded incorrectly.}
\label{yufig2}
\end{figure}

In this geometric representation, the effect of an additive noise is
to add a noise vector to the signal vector.  The effect of a power
constraint $P$ on $X^{(in)}(t)$ is to have it lie inside a $D$-dimensional
sphere of radius $\sqrt{P}$ in the whole space $R^D$, the Euclidean
space of dimension $D$.  For white Gaussian processes, the coefficients
of its expansion in any orthonormal basis are independent Gaussian
random variables, with variances all given by the same quantity, the
average power of the process [6-8,13]. The receiver looks at
the received point A in the signal space, and picks the nearest signal
point in Euclidean distance to A for minimizing the error probability
assuming equiprobable messages.  The situation is illustrated in
Fig.~\ref{yufig2}.

Thus, in time $T$ there are $2TW$ independent Gaussian
amplitudes from (\ref{y20}), and from (\ref{y19}) the total number of well
distinguished signals is
\beq
M = \left[ k \sqrt{\frac{P+N}{N}} \right] ^{2TW} . \label{y21}
\eeq
The number of bits per second is, from (\ref{y9}),
\beq
\frac{\log _2 M}{T} = W\log _2 k^2 \frac{P+N}{N} . \label{y22}
\eeq
The capacity formula (\ref{y15}) for AWGN channel follows from (\ref{y22})
with $k=1$. It comes about more precisely as follows.  As a result of the
statistical regularity mentioned above, for large $T$ the
signals $X^{(in)}(t)$ must almost all lie on a sphere of radius
$\sqrt{2TWP}$, signal plus noise $X^{(out)}(t)$ on a sphere of radius
$\sqrt{2TW(P+N)}$, with noise $n(t)$ on a sphere of radius
$\sqrt{2TWN}$ centered at the original signal point.  Note that the
Euclidean structure of the signal space, which is absent in the
general discrete case, is crucial here --- the different coordinate
values of, say, $n(t)$ in the $D$-dimensional space are Gaussian
distributed yielding the given average value $DN$ for the norm $\int_T
n^2(t) dt$ of the $n(t)$ vector, so that $n(t)$ is on a sphere of
radius $\sqrt{DN}$ around the source signal point.  For arbitrarily
small error probability, one would want the noise spheres around
different signal points to overlap arbitrarily little.  A
``sphere-packing'' argument (see (\ref{y35}) below also) then readily
establishes the converse to the coding theorem for (\ref{y15}), namely that
it is impossible to transmit with arbitrarily small error probability
at rates above $C$.  For the positive statement that it is indeed
possible to so transmit at rates below $C$, a ``random coding'' argument
is required which in fact establishes the following amazing result: if
the signals are selected at random, with probability one the resulting
error probability is arbitrarily small.  The dichotomy at $C$, for all
rates $R$ below $C$ the block error $P_e \rightarrow 0$ for almost all
long codes ($n \rightarrow \infty$) while for rates above $C$, $P_e
\rightarrow 1$ for almost all long codes, is exactly like a phase transition.
In practice, it turns out that long codes or signal sets that have
enough structure to be readily described, encoded and decoded, do not
approach capacity although the situation seems to be changing very
recently.  For more details of the above description see [6] and [13].

Besides communication of information, the problem of estimating a
continuous entity is also of prime concern in this chapter.  Consider
a Gaussian random variable $U$ with zero mean (or normalize it away)
and variance $\sigma ^2$, which is received in the form $Au$ in
Gaussian noise with a possible gain or loss $A$, i.e.,
\beq
X^{(out)} = Au + n . \label{y23}
\eeq
This may arise in linear modulation, or in the estimation of $U$ in
any experiment.  (The word ``detection'' is commonly reserved for the
``estimation'' of a discrete $U$.)  If the mean-square error
$\ovr{\epsilon^2}$ between the estimate $V=\hat{u}(X^{(out)})$ and $U$ is to be
minimized, the best estimate is given by [6,8] the conditional
mean $E[U|X^{(out)}]$,
\beq
\hat{U}_{MMSE} (X^{(out)}) = \frac{X^{(out)}/A}{1+N/\sigma ^2 A^2} , \label{y24}
\eeq
where $N$ is the noise variance, with resulting
\beq
\ovr{\epsilon^2} = \frac{\sigma ^2}{1+\sigma ^2 A^2/N} \equiv \sigma ^2
(1+\frac{S}{N})^{-1} \label{y25}
\eeq
in terms of the a priori variance $\sigma ^2$ and a signal-to-noise
ratio $S/N$.

In the physics literature on quantum information and its applications, 
the criterion of mutual information is often used in place of detection
or estimation error in situations (such as cryptographic eavesdropping) 
for which no coding is possible.  Depending on the problem, the best 
possible outcome of such use would be a bound rather than the desired 
performance criterion.

\subsection{Communication versus measurement}\label{ysec.2.3}

The estimation/detection problem clearly parallels the problem of
producing an estimate of a desired quantity $u$ from the measured
data $X^{(out)}$ in a physical experiment. More generally, in an estimation
problem one is given a fixed statistical specification $X^{(out)}(t,u)$ and
forms in general a nonlinear estimate $\hat{u}(X^{(out)})$ so that a cost
function $C(\hat{U}, U)$ is minimized. For mean-square error,
\beq
C(\hat{U}, U) = \ovr{\epsilon ^2} = \mbox{E}[|U-\hat{U}|^2] , \label{y26}
\eeq
where the expectation E is taken over all the random quantities
involved.  In a communication situation, the channel is a statistical
transformation $F$ on the input $X^{(in)}(t,u)$
\beq
X^{(out)}(t,u) = F[X^{(in)}(t,u)] , \label{y27}
\eeq
which reads, for an additive noise channel,
\beq
X^{(out)}(t,u) = X^{(in)}(t,u) + n(t) \label{y28}
\eeq
for which one can control $X^{(in)}(t,u)$ subject to the system constraints.
In contrast to the estimation case, a direct optimization approach to
a communication problem with joint transmitter/receiver optimization
has never been developed in a useful way.  Instead of asking for the
optimum system for a fixed time duration $T$, $T$ itself is floated as a
design parameter in the development of channel encoding-decoding
design subject to Shannon's coding theorems.

It can be seen that
a {\em physical measurement} is generally {\em not} just an estimation
problem, because $X^{(out)}(t,u)$ can be influenced to some extent through the
choice of the physical measurement process, although perhaps not as
much as controlling $X^{(in)}(t,u)$ in (\ref{y27}). In particular, it is a major
part of the measurement system design to find an appropriate {\em
physical variable} $X^{(in)}$ to couple to the desired information parameter
$u$ to form $X^{(in)}(t,u)$ for information extraction after the corruption
of $X^{(in)}(t,u)$ to $X^{(out)}(t,u)$ by the ``channel'' is taken into account.
However, there is usually no question of data rate in a measurement.
Thus, physical measurements, which are of prime concern to us, are
described somewhere between communication and detection/estimation.
This situation already obtains in classical measurements, and in cases
of fixed quantum states and quantum measurements.  It becomes more so
in quantum communications where the quantum states and quantum
measurements can be freely chosen.  As developed in Section~\ref{ysec.4},
the feasibility of choosing quantum states moves a physical measurement
problem away from being a pure estimation problem to becoming more
like a communication problem, although it never fully becomes a
standard communication problem.

\section{Quantum communication} \label{ysec.3}

\subsection{Quantum Versus Classical Communication} \label{ysec.3.1}

By ``quantum communication'' we mean more than the study of quantum
effects in communication systems involving classical information
transfer. Specifically, in quantum communications we are concerned
with the system performance under a variety of different quantum
measurements and quantum states. Referring to Fig.~\ref{yufig1},
the statistical
specification of the channel plus transmitter, e.g., is given by a
conditional probability $p(X^{(out)}|u)$.  The classical variable
$X^{(out)}$ may well
be of quantum origin, say it is the eigenvalue of a quantum
observable obtained in a measurement.  However, as far as the analysis
of this system is concerned, the fact that $p(X^{(out)}|u)$ arises from
quantum mechanics makes no difference, and it would proceed just like
a classical communication system.  If this $p(X^{(out)}|u)$ arises from a
quantum state $\rho(u)$ and a quantum measurement of a selfadjoint
$\hat{X}^{(out)}$ with eigenstates $|X^{(out)}\rang$, $p(X^{(out)}|u) = \lang
X^{(out)}|\rho (u)|X^{(out)}\rang$, one may well ask whether other possible
choices of $\rho (u)$ and observable with resulting different $p(X^{(out)}|u)$
may lead to
better performance.  These additional freedoms of quantum measurement
and quantum state selection are absent in a classical communication
system. They constitute the new content of {\em quantum communication}.

\begin{figure}
\includegraphics[width=\textwidth]{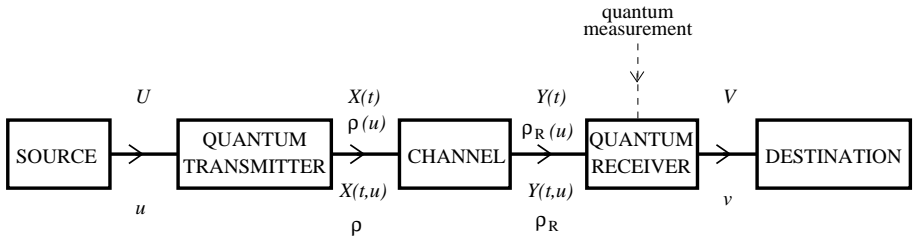}
%%\centering\psfig{file=./yue3.eps,width=.9\textwidth}
\caption{Schematic representation of a quantum communication
system: the message dependence generally enters through the state for
the field, but can be put in the field operator itself in some
instances.}
\label{yufig3}
\end{figure}

A general quantum communication system is depicted schematically in
Fig.~\ref{yufig3}. The channel input and output signals $\hat{X}^{(in)}(t)$ and
$\hat{X}^{(out)}(t)$ are now field operators in quantum states $\rho(u)$
and $\rho_R(u)$.

Generally, as indicated by the Dimensionality
Theorem (\ref{y20}), a finite number of modes each with two degrees of
freedom, such as an optical mode with two quandratures, would suffice
so that the density operators $\rho$ and $\rho_R$ are well-defined.  A
specific classical statistical characterization of the system would
result upon a choice of quantum measurement at the receiver.  The most
general characterization of a quantum measurement is the so-called
``completely positive operation measure'' with the corresponding
measurement statistics given by a positive operator-valued measure
(POM) [16].  Let $\hat{O}(X^{(out)})$ be a POM on $\hat{X}^{(out)}(t)$, the
channel output, thus
\beq
\sum_{X^{(out)}} \hat{O}(X^{(out)}) = I \qquad \mbox{or} \qquad
\int \hat{O}(X^{(out)}) dX^{(out)} = I \label{y29}
\eeq
and each $\hat{O}(X^{(out)})$ is a nonnegative selfadjoint operator, with
$\hat{O}(X^{(out)}) = |X^{(out)}\rang \lang X^{(out)}|$ for orthogonal
$|X^{(out)}\rang$ in the case
of a selfadjoint observable. The statistics are given by
\beq
p(X^{(out)}|u) = tr\rho_R(u)\hat{O}(X^{(out)}) . \label{y30}
\eeq
The output state $\rho_R(u)$ is determined by the channel action on
the input state $\rho(u)$.  The additional quantum ``freedoms'' in
quantum communication consist in the selection of $\hat{O}(X^{(out)})$ and
$\rho(u)$. Note that it may be more convenient, as in the case of
frequency modulation, to enter the information variable $u$ in
$\hat{X}^{(in)}(t)$ in parallel with the classical case, and specify the
transmitter in terms of $\hat{X}^{(in)}(t,u)$ and $\rho$ rather than
$\hat{X}^{(in)}(t)$ and $\rho(u)$. In this formulation, the quantum state
$\rho$ and the classical modulation process are decoupled. However,
if the information $u$ enters through the quadratures, it would be
necessary to use $\rho(u)$, for which the quantum state selection and
modulation selection are tied together.

Historically, the serious study of optical communication began
immediately after the laser was experimentally realized, for which
quantum effects are clearly important as $\hbar \omega/k \sim 10^{4}$
$K$.  While the evaluation of system performance went on for
coherent states and the three standard measurements: direct, homodyne,
and heterodyne detections, quantum communication theory in our sense
was also developed. Forney [17] and Gordon [18] proposed the
entropy bound for information transfer with a fixed set of states
valid for arbitrary measurement, to be discussed in section~\ref{ysec.3.3}.
Helstrom [19] studied the quantum measurement optimization
problems in the spirit of classical detection/estimation theory, which
were further developed by Holevo [20] and Yuen [21].  Each of the above three
standard measurements corresponds, respectively, to the quantum
measurement of photon number, single field quadrature, and joint
quadratures described by a POM but not a selfadjoint
operator [22]. In such work, which actually has many applications
in physics [23] but will not be further discussed in this chapter,
the states are fixed and the quantum measurement is selected so that
the resulting classical statistical system leads to the best possible
performance compared to other measurements.  The issue of quantum channel
representation was treated [4,24,25] and the possibility of receiver 
state control is suggested [2,4,25,26,27].  The general problem of
transmitter quantum state selection 
was considered by Yuen [3,4], leading to the
development of TCS as indicated in Section~\ref{ysec.1}.  The question of optimal state
influence on channel capacity was also implicit in connection with the
application of the entropy bound, indicating that number states and
photon counting are best for free boson fields [4,28].  For recent advances
in determining the capacities and error exponents of various quantum
channels by Holevo and the Hirota group, see [29-31].  For other advances
including work on quantum tomography by the D'Ariano group, 
see [32-34].  For applications of squeezed states to quantum cryptography, 
see [35, 36].

\subsection{Mutual information} \label{ysec.3.2}

The capacities, or mutual informations maximized over the input
distributions, for various boson channels are discussed extensively in
[37].  Here I would like to focus on five capacities for the
narrowband free boson channel under an average power constraint:
number state and photon counting, TCS and homodyning, coherent state
and the three standard measurements.  Hopefully, it would become clear
within this and the next subsection that they are the most important
cases capturing the essence of the situation.

For the free electromagnetic field at optical frequencies, all the
current or forseeable future systems are narrowband, i.e., the
available bandwidth is only a small fraction of the center
frequency.  Due to various facts of nature, it would be extremely
difficult and inefficient to utilize photons at higher frequencies,
say X-rays, in a communication situation.  Thus, there is no practical
significance in studying wideband photonic channels.  The constraint
of average power can be separated into two parts: average with respect
to the statistics of the information variable $U$ and average with
respect to the quantum nature of the state $\rho$. In either case a
peak power (or energy or power spectral density) constraint can also
be applied.  In the case of classical signals the peak power
constraint is indeed quite meaningful and realistic, but is often hard
to handle mathematically and usually avoided.  In the case of
quantum states, a peak energy constraint would cut off the Hilbert
space of states $\cal H$ at a maximum number state eigenvalue $n_m$ so
that $\cal H$ becomes finite-dimensional.  This, however, is
unrealistic or at least hard to handle in so far as one considers 
a coherent state $|\alpha
\rang$, which has components in all $|n\rang$, to be realizable.  Some
discussion on this issue is given in [10], although
in its full scope it is a complicated and profound issue.  Here I
would advocate, if only on the ground of mathematical convenience,
that energy constraint is to be applied to the quantum state average
$tr\rho a^{\dagger}a$, and not to yield an $n_m$.

\begin{figure}
\includegraphics[width=.8\textwidth]{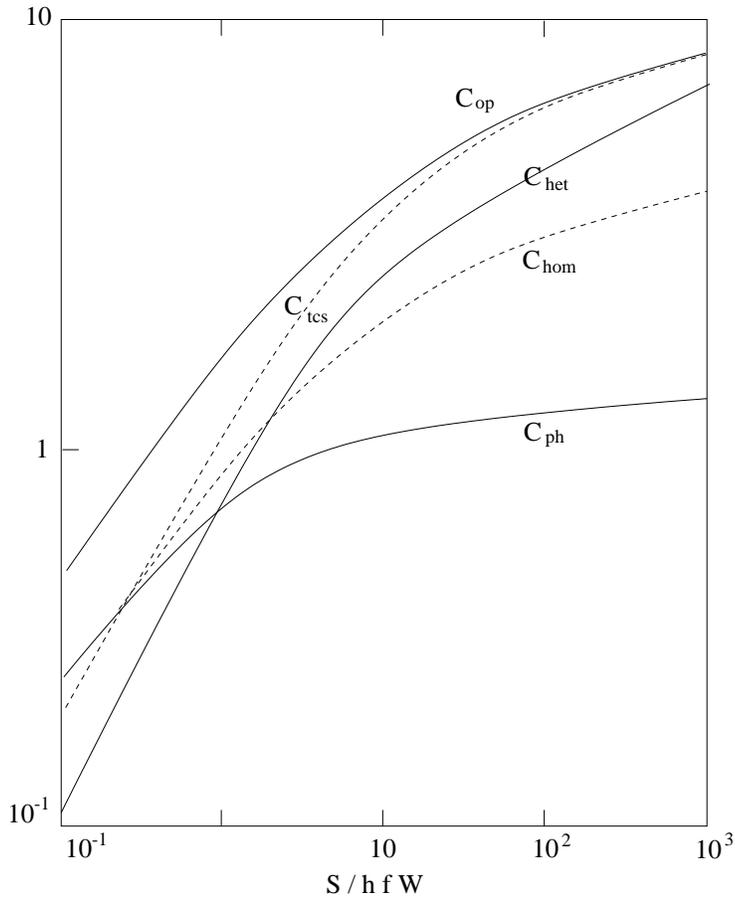}
\caption{Comparison of capacities in bits per second for $f_0
\sim 5.7 \times 10^{14}$ Hz and $W \sim 10^{14}$Hz.}
\label{yufig4}
\end{figure}

Let $P = hf_0 WS$ be the available signal power of a narrowband
channel of center frequency $f_0$ and photon numbers $S$ per mode.
The photon number capacity is [28,37,38]
\beq
C_{op} = W[(S+1)\log(S+1)-S \log S] . \label{y31}
\eeq
For TCS with homodyne detection [4,37],
\beq
C_{TCS} = W \log (1+2S) . \label{y32}
\eeq
If both quadratures of the TCS are utilized under the same power
constraint with optimized TCS-heterodyne [22] or
joint quadrature measurement, it can be shown from the Kuhn-Tucker
optimizality conditions of nonlinear programming that as $S$ is
increased from 1 the optimum capacity is indeed achieved through
utilization of only one quadrature.  The coherent state heterodyne and
homodyne capacities are
\begin{eqnarray}
C_{het} & = & W \log (1+S) , \label{y33}\\
C_{hom} & = & \frac{W}{2} \log (1+4S) . \label{y34}
\end{eqnarray}
Equations (\ref{y32})-(\ref{y34}) are easily derived from (\ref{y15}) because
the corresponding channels are AWGN ones --- the fluctuations in a TCS or
a coherent state, which is merely a classical amplitude superposed on
vacuum, are behaving as independent additive Gaussian noises, and are
white noises under the narrowband assumption.  The coherent-state
photon counting capacity $C_{ph}$ does not have a simple closed form but is
readily computed numerically [39].  These five
capacities are compared numerically in Fig.~\ref{yufig4} reproduced
from~[4], for a fairly wide bandwidth.  It may be
observed that $C_{TCS}$ is always larger than $C_{het}$ and $C_{hom}$,
and is also larger than $C_{ph}$ in the case of more than a fraction of a 
photon per mode.

However, the difference between $C_{TCS}$ and $C_{het}, C_{hom}$ is
not big.  Indeed, $C_{TCS}$ is less than twice $C_{hom}$ although the
SNR of TCS is the square of that for coherent states.  Because the
data rate for a mode goes as log(1+SNR) from (\ref{y15}), the square in SNR
becomes less than a multiplicative factor of~2.  The difference
between $C_{TCS}$ and $C_{het}$ is even less, (\ref{y32}) is equivalent to
doubling the signal power in $C_{het}$ with the same bandwidth.  The
underlying reason can be understood as follows.  In the geometric
representation of signal and noise sketched in section~\ref{ysec.2.2},
it can be
seen that the effect of noise is to move a given signal point away
from its position.  If the noise is big enough, it would move it
closer to another signal point B as compared to the original point A,
and the optimum receiver would decide it is this other signal B that
was transmitted, hence making an error, as illustrated in
Fig.~\ref{yufig2}.
Thus, a good system would have the signal points as far apart as
possible from the viewpoint of errors, and have as many signal points
as possible from the viewpoint of data rate, two conflicting goals.
For a fixed dimension $D \sim 2WT$, a larger power $P$ yields a larger
sphere and the same number of M signal points can be placed further
apart inside the sphere, leading to a smaller error for a fixed noise
power $N$.  Increasing $W$, however, is more beneficial than
increasing $P$, thus $C_{het} > C_{hom}$ as $W$ increases, even though
$C_{hom}$ has a bigger SNR.  To see the role of $W$ versus $P$, recall
the discussion around (\ref{y21}) and (\ref{y22}) that one wants the noise
spheres around different signal points to be almost nonoverlapping to
yield small error probability.  As a result of this ``sphere packing,''
the number of well distinguished signals is roughly the ratio of the
signal plus noise volume to the noise volume.  The volume $V_D(r)$ of a
$D$-dimensional sphere of radius $r$ is $B_Dr^D$ for a $D$-dependent
constant $B_D$, which implies
\beq
V_D (\sqrt{D(P+N})/V_D(\sqrt{DN}) = (1 + \frac{P}{N})^D \label{y35}
\eeq
since the radii of the signal plus noise and noise spheres is
$\sqrt{2TW(P+N)}$ and $\sqrt{2TWN}$ respectively.  The quantity 
(\ref{y35}) grows exponentially in $D$ or $W$ but only to a fixed power in
$P$.  This more important role of $W$ versus $P$ clearly manifests in
(\ref{y15}) and (\ref{y16}).

Having understood why the apparent large gain in SNR given by (\ref{y8}) for
TCS leads only to a small gain in capacity, the question becomes
whether TCS would be significant in improving optical communications
compared to coherent states.  This rest of this section~\ref{ysec.3} is devoted
to a detailed examination of this issue.  We may first observe that
complicated coding, especially the decoding process, is required to
approach capacity given by any of (\ref{y32})-(\ref{y34}).  If one looks at the
error behavior of information transfer under specific simple signaling
scheme, e.g., the antipodal signals discussed in [5],
the full SNR square advantage may appear.  That is, more restricted
``capacities'' than (\ref{y32})-(\ref{y34}) may show a large advantage with TCS.
In Section~\ref{ysec.3.3} we will see that the number state capacity $C_{op}$,
which is so close to $C_{TCS}$, is actually the optimum rate for any
states and measurements subject to the average power constraint.  This
capacity $C_{op}$ can be obtained {\em without} the need for
complicated decoding because the ideal number state channel is
noiseless --- there is no need to use long sequences to yield
statistical regularity. Thus, the use of number states can be
considered as an {\em alternative to channel coding}.  Number states,
as intensity squeezed states, have a lot of similarity to TCS in
regard to their physical generation and propagation characteristics.
Unfortunately, the use of such nonclassical states as information
sources would {\em not} be advisable in practical communication
systems.  In addition to various problems of a more practical nature,
such as phase coherence for TCS and good detectors for number states,
the inevitable presence of significant loss would wipe out the
advantage of nonclassical states.  This issue will be treated in
section~\ref{ysec.3.4} after the following discussion of the entropy bound that
established the optimality of $C_{op}$ given by (\ref{y31}).

\subsection{The entropy bound} \label{ysec.3.3}

Given a fixed set of density operators $\rho_{\lambda}$ dependent on a
discrete or continuous random variable $\Lambda$ with probability
(density) $p(\lambda)$, define
\beq
\ovr{\rho} \equiv \sum_{\lambda} p(\lambda) \rho_{\lambda}
\qquad \mbox{or} \quad \int p(\lambda) \rho_{\lambda} d\lambda .
\label{y36}
\eeq
Let $S(\rho) \equiv -tr\rho \log \rho$ be the Von Neumann
entropy of
$\rho$, and let $\hat{O}(X^{(out)})$ be the POM giving the measurement
probability. Then the mutual information between $\lambda$ and $X^{(out)}$
is bounded by
\beq
I(\Lambda ; \mbox{Y}) \leq S(\ovr{\rho}) - \ovr{S} (\rho_{\lambda})
\label{y37}
\eeq
\beq
\ovr{S} (\rho_{\lambda}) = \sum_{\lambda} p(\lambda) S(\rho_{\lambda})
\qquad \mbox{or} \qquad \int p(\lambda)S(\rho_{\lambda})d\lambda .
\label{y38}
\eeq
This entropy bound (\ref{y37}), first given by Forney [17] and
Gordon [18], was proved for finite discrete $\Lambda$ and finite
dimensional Hilber space~$\cal H$ by Zador~[40] and independently
by Holevo [41], and general $\Lambda$ and infinite dimensional
$\cal H$ by Ozawa [38].  The long complicated history of this
bound is outlined in~[10].  Recently, the inequality in (\ref{y37}) is
shown to be achievable if the measurement can be made over a long
sequence of states instead of just symbol by symbol in the sequence
[42,43].  However, while this may be considered to establish
the capacity of a quantum channel defined by the mapping~$\lambda
\mapsto \rho_{\lambda}$, such a specification of a quantum channel is
neither general nor practical.  The main reason is that there is no
way to tell whether the particular map $\lambda \mapsto
\rho_{\lambda}$ is optimum under the constraint of the problem.  As we
have emphasized in section~\ref{ysec.3.1}, a general quantum communication
problem involves {\em both} the choice of states and measurements.
Under an average energy constraint for a single mode,
\beq
\sum_{\lambda} p(\lambda)tr \rho_{\lambda}a^{\dagger} a \leq S \label{y39}
\eeq
one cannot tell a priori what the optimal $\lambda \mapsto \rho_{\lambda}$ should
be, even if one assumes all $\rho_{\lambda}$ are coherent states.

On the other hand, the bound (\ref{y37}) in its full generality readily
shows~[38] that for a single boson mode under (\ref{y39}),
the maximum $I(\Lambda ; \mbox{Y})$ is achieved by taking $\lambda$
and $X^{(out)}$
as a nonegative integer, with number states $\rho_{\lambda = n} = |n \rang \lang n
|$ and the $I$ value given by (\ref{y31}) for $W = 1$.  The wideband capacity
can be similarly derived~[38].  Note that apart from
showing that more general processing such as feedback would not
increase $I$, this joint optimization over state/measurement and
modulation (the map $\lambda \mapsto \rho_{\lambda}$) demodulation
(the map $X^{(out)} \mapsto \lambda$) does establish that (\ref{y31})
is the {\em ultimate} quantum limit on the possible rate of information
transfer for a boson mode of average energy $S$.

\subsection{Effect of loss}\label{ysec.3.4}

The effect of linear loss on a mode can be represented as a
transformation on the modal photon annihilation
operator [2,4,44-46],
\beq
b = \eta^{\frac{1}{2}} a + (1-\eta )^{\frac{1}{2}} d , \label{y40}
\eeq
where $\eta$ is the transmittance and the $d$-mode is in vacuum.  Note
that the effect of quantum efficiency on a detector can be so
represented as well.  It follows immediately from (\ref{y40}) that the
resulting quadrature fluctuation in $b$ has a floor level $(1-\eta)/4$,
which is essentially the coherent state noise level for $\eta \ll 1$.
Similarly for a number state, the $b$-mode photon number fluctuation
\beq
\lang \Delta N^2_b \rang = \eta^2 \lang \Delta N^2_a \rang + \eta (1 - \eta)
\lang N_a \rang \label{y41}
\eeq
contains a partition noise equal to the mean $\lang N_b\rang = \eta
\lang N_a \rang$ for $\eta \ll 1$, washing away the sub-Poissonian
character of the $a$-mode.  Generally, one can readily show from~(\ref{y40})
that the state $\rho_b$ is very close to a coherent state of mean
$\eta^{1/2} \lang a\rang$ for large loss, thus any
nonclassical state becomes essentially classical.

The implication of this fact on the utility of nonclassical states is
profound, especially in engineering applications where significant
loss is usually present, e.g., in fiber optic communications.  Unless
a special environment is created [4] to compensate for
the squeezing or nonclassical effect in the presence of loss, there is
no way to keep a nonclassical state at the reception end.  While this
is possible in principle, it seems that is not worth the
trouble.  Even in scientific experiments or in the process of
nonclassical state generation, loss places a severe limit on the
amount of squeezing obtainable. The {\em sensitivity} of nonclassical
states to loss and interference would place strenuous requirements on
all the system components, making any such system extremely difficult
to implement. To me, a similar kind of argument leads
to a similar implication in the field of quantum information.

Given the close value of $C_{TCS}$ to $C_{het}$ in (\ref{y32})-(\ref{y33})
in the absence of loss, it should be clear that there is hardly any
advantage left in the presence of loss.  While the ultimate quantum
capacity $C_{\ell}$ of a lossy channel is not known, an upper bound on
$C_{\ell}$ can be easily derived.  Under an average energy constraint
$S$ and loss $\eta$, equation (\ref{y31}) for $C_{op}$ with $S$ replaced by
$\eta S$ would provide a bound on $C_{\ell}$.  The gap between
$C_{op}$ and $C_{het}$ with $\eta S$ is the largest gain, probably not
actually achievable, that one can possibly obtain with nonclassical
states in a lossy channel.  The smallness of this gap, as seen from
Fig.~\ref{yufig4}, shows that there is little significance in pursuing
quantum communications with nonclassical sources in practice, a conclusion
I drew over twenty years ago.

\subsection{Quantum amplifiers and duplicators}\label{ysec.3.5}

Not all is lost, however.  As to be presently explained, the use of
novel quantum amplifiers and related devices on coherent state sources
can lead to a number of significant communication applications not
possible with the usual phase-insensitive linear amplifier (PIA).  A
characteristic feature of these novel devices is that their outputs
are often nonclassical states for coherent state inputs, even though
it is not the nonclassical nature of these states that is relevant in
the application.

Corresponding to the three standard quantum measurements are three
quantum amplifiers, the photon number amplifier (PNA), the
phase-sensitive linear amplifier (PSA), and the PIA.  If $b$ and $a$ are
the output and input modal photon annihilation operator of the
amplifer, these three amplifiers can be represented
as [45-48], with a power gain $G > 1$,
\begin{eqnarray}
 \mbox{PIA} \hspace*{.2in} b & = & G^{1/2} a + (G-1)^{1/2} v^{\dagger},
 \hspace*{.2in} [v,v^{\dagger}] = 1 \label{y42}\\
 \mbox{PSA} \hspace*{.2in} b_1 &  = & G^{1/2} a_1 , \hspace*{.2in}
 b_2 = G^{-1/2} a_2  \label{y43}\\
 \mbox{PNA} \hspace*{.2in} b^{\dagger}b & = & Ga^{\dagger}a,
 \hspace*{.2in} \mbox{G integer} \label{y44}
\end{eqnarray}
A fourth quantum phase
amplifier [49,50] is
\beq
\mbox{QPA} \hspace*{.2in} e_+ = e_+^G , \hspace*{.1in} e_+ \equiv
(a^{\dagger} a+1)^{\frac{1}{2}}a^{\dagger}, \label{y45}
\eeq
which is related to the ideal phase measurement [19,21,51]
described by a POM involving the Susskind-Glogower states and
corresponding phase-coherent states [51].

\begin{center}

{\bf Table 1}

\vspace{12pt}

\begin{tabular}{c|c|c|c} 
{\bf DETECTION} & {\bf AMPLIFIER} & {\bf STATES} & {\bf DUPLICATORS} \\ \hline \hline
heterodyne & PIA & CS & BQD \\ \hline
homodyne & PSA & TCS & SQD \\ \hline
direct & PNA & NS & PND \\ \hline
phase(ideal) & QPA & PCS & QPD \\ 

\end{tabular}

\end{center}

\noindent (The column on states merely emphasizes that the nature 
of these states would be preserved only by the corresponding
amplifier, not that the amplifier is noiseless only for those
states.)

\begin{center}

{\bf Abbreviations for Table and Text}

\vspace*{12pt}

\begin{tabular}{ll}
{\bf CS} & coherent state\\
{\bf NS} & number state\\
{\bf TCS} & two-photon coherent state\\
{\bf PCS} & phase coherent state\\
{\bf PIA} & phase-insensiive linear amplifier\\
{\bf PSA} & phase-sensitive linear amplifier\\
{\bf PNA} & photon-number amplifier\\
{\bf QPA} & quantum phase amplifier\\
{\bf PND} & photon number duplicator\\
{\bf POA} & photon on-off amplifier\\
{\bf QND} & quantum nondemolition measurements\\
{\bf POM} & positive operator-valued measure\\
{\bf SQL} & standard quantum limit
\end{tabular}

\end{center}

In (\ref{y42}) and (\ref{y44}), the photon operator $b$ has to
be defined on two modes.  For a fuller discussion of these amplifiers,
see [48] and [49] which also contains an extensive treatment of
duplicators to be discussed later in this section.  The main point
about (\ref{y42})-(\ref{y44}) is that the amplifier output of each is, for the
corresponding measurement, a perfect ``noiseless'' scaled (amplified)
version of the input for {\em arbitrary} input state, i.e., they are
noiseless amplifiers for the corresponding detection scheme.  Thus,
the often found statement that quantum amplifier necessarily
introduces noise, say in the sense of having a noise figure F $\equiv$
SNR$_a$/SNR$_b > 1$, is {\em wrong}.  As summarized in Table 1, if the
proper amplifier matching the measurement is used there is no
additional noise ideally, similar to the classical case.  
All the noise then arises {\em
inherently} from the quantum nature of the input.  (This is also true
in both balanced and unbalanced homodyne/heterodyne detection for 
which the effective amplifier, the local oscillator, introduces no noise
in the high gain limit.  See [52].  It is a pervasive misconception that 
the noise in homodyne/heterodyne detection is local-oscillator shot noise.)
Without going into
a detailed exposition, this is actually clear intuitively from the
basic principles of quantum physics.  When you fix a measurement, the
situation is classical for any given state as discussed in
Section~\ref{ysec.3.1} on quantum vs. classical communication, {\em in the
sense that} a fixed
probabilistic description is obtained.  The situation is a little more
subtle in the case of POM rather than selfadjoint operator, but can be
understood by analyzing the POM as commuting selfadjoint operators
measurement on an extended space which can always be done [20].

The generation mechanism of PSA is identical to quadrature squeezing, which,
being piecewise linear, is not exactly a nonlinear effect.  On the
other hand, PNA, QPA and the duplicators involve truly nonlinear
quantum effects~[47-50] which would not be discussed here.
None of these new quantum devices except PSA has been successfully
demonstrated experimentally in a useful manner.

At this point, I would like to address a confusing point about the
capability of amplifiers.  It is often stated that an amplifier at the receiver
could improve the receiver performance.  The optimum receiver
performance is determined by the specification $\{ \hat{X}^{(out)} (t),
\rho (u) \}$ in Fig.~\ref{yufig3}.  Nothing, and certainly no
amplifier, can ever
improve that as a matter of tautology.  What can be improved is a
specific receiver structure that does not lead to the optimum
performance. In such a case, the use of an amplifier or some other
device may improve the suboptimum receiver performance.  This point is
related to, but different from, the so-called data processing
theorem [7] in information theory which shows that no
processing can increase the information transfer over a channel.

The above amplifiers can be used as pre-amplifiers to suppress
subsequent receiver noise in the corresponding detections, in either
engineering or scientific applications.  They can also be used to
advantage [53] in the attempt to create a transparent
optical local area network.  For such a purpose, however, the
duplicators [46-48,54] would be perfect.  A
photon number duplicator (PND) is a device with one input $a$ in state
$\rho_a$ and two outputs $b,c$ such that each of the output photon
counting statistics is the same as that of the input
\beq
\lang n| \rho_a |n \rang = \lang n |\rho_b |n \rang = \lang n|
\rho_c |n \rang . \label{y46}
\eeq
Typically, the output photon counts for the $b$ and $c$ modes are
perfectly correlated, thus PND also provides a perfect realization of
a photon number quantum nondemolition measurement (QND) with only a
finite energy [47].  Single and double quadrature
duplicators can be similarly described.

\begin{figure}
\includegraphics[width=0.8\textwidth]{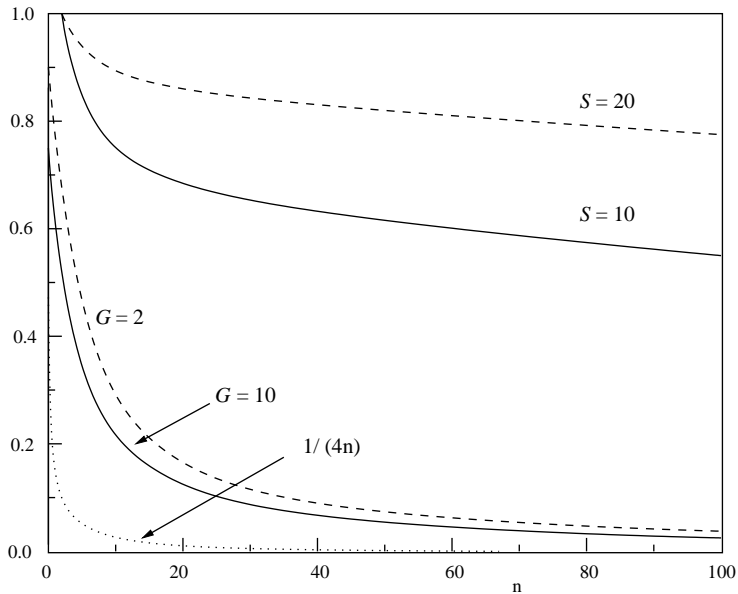}
\caption{ Comparison of error exponents $S^{-1}$ln 2$P_e$ as a
function of stages $n$ --- the PIA line exponent is
independent of $S$ and $G$, the PNA exponent is independent of $S$ and
the POA exponent is independent of $G$.}
\label{yufig5}
\end{figure}

The amplifiers can be used as line amplifiers in long distance optical
fiber communications.  For example, the use of PSA not only improves
the SNR by a factor of 2 for coherent state sources in a long
amplifier chain, it also significantly reduces the Gordon-Haus soliton
timing error [55].  Considerable experimental progress [56]
has been made on such possible application, but the required phase
coherence renders it impractical.  For on-off signals, the use of PNA
leads to the following error probability
\beq
P_e = \frac{1}{2} \exp \{ -S[1-f_n(G)] \} , \label{y47}
\eeq
where the functions $f_n(G)$ obey the recurrence relation
\beq
f_{n+1}(G) = (1-G^{-1})^G [1+(G-1)^{-1}f_n(G)]^G \label{y48}
\eeq
with $f_0(G) = 0$.  Equations (\ref{y47})-(\ref{y48}) apply to a chain of $n$
amplifiers of gain $G$ and loss $G^{-1}$ between two adjacent
amplifiers, assuming direct detection.  In Fig.~\ref{yufig5}, this error
exponent $1-f_n(G)$ is compared with that of the PIA line,
$\frac{1}{4n}$, obtained under the Gaussian approximation for direct
detection.

As can be seen in the figure, even more improvement, in fact the
optimum improvement, is obtained with the use of a photon on-off
amplifier [57] (POA) tailored for the situation.  In
the state description, a POA acts on two modes but for the input mode
it reads
\begin{eqnarray}
|0\rang & \mapsto & |0\rang \nonumber \\
\mbox{POA} \hspace*{.2in} |1\rang & \mapsto & |\alpha \rang \label{y49}\\
        & \vdots & \nonumber \\
|n\rang & \mapsto & |\alpha\rang \nonumber \\
        & \vdots & \nonumber
\end{eqnarray}
where $|\alpha \rang , |0\rang$ are the two on-off coherent states, $S
= |\alpha |^2$.  The resulting error probability is
\beq
P_e = \frac{1}{2}[1-(1-e^{-S})^n] , \label{y50}
\eeq
which is the same as that obtained by a repeater, i.e., by direct
detection and retransmission at each of the n stages.  In general, it
is possible to write down a {\em perfect} quantum amplifier for any given
signaling and detection scheme which performs as well as a repeater,
although the actual installation of POA or any such amplifiers in a
long line would entail the loss of flexibility, as compared to PNA,
for adapting to other signaling schemes.

Quantum amplifiers are also useful in quantum cryptography [48].
A major problem of the quantum cryptographic schemes is that
they cannot be amplified to compensate for the loss without disrupting
the operation of the scheme.  In [58] a new quantum cryptographic
scheme is introduced that allows amplification, which greatly extends
the distance over a fiber for which the scheme works.  

\section{Ultimate limit on measurement accuracy} \label{ysec.4}

\subsection{Measurement System and Ultimate Performance}
\label{ysec.4.1}

In this section~\ref{ysec.4}, the question is addressed on the ultimate,
quantum as well as classical, limit on the measurement accuracy obtainable
with various measurement systems.  The optimum performance ideally
achievable with a measurement system is of course an important piece
of design information, but more importantly I would like to assess the
potential of such systems, and ways to realize them in principle, in
order to explore the feasibility of developing ultrahigh precision
measurement systems important in many applications, especially
scientific ones.  My approach [58] is based on the
communication characterization of measurement discussed so far,
especially in section~\ref{ysec.2.3}, while adopting quantum and classical
communication theory to provide the answers.  Since the correspondence
between communication and measurement is not exactly isomorphic, we
will find that it is possible to obtain limits on the measurement
accuracy, but not always possible to be assured that those limits are
attainable.  Indeed, even if the correspondence is perfect, there are
still additional questions, such as what systems are actually available, that
would resist a complete mathematical characterization in the
forseeable future.  Nevertheless, as to be discussed presently, some
of the results obtained are somewhat surprising, and also promising.
In the next section~\ref{ysec.4.2} the rate distortion limit in classical
communication theory will be explained, and in section~\ref{ysec.4.3} the
corresponding quantum limits will be presented.  Here I would like to
first highlight the results and their implications.

The final error in a measurement system may depend, even in principle
excluding nonideal environmental perturbations, on more than a single
source or variety.  For example, in the detection of very weak
gravitational radiation by a Michelson interferometer, the radiation
pressure error needs to be added to the photon detection error to form
the total error.  The application of squeezed states in this situation
is treated elsewhere in this book and would not be discussed.  Here,
the general theory would be illustrated only with a measurement medium
or channel that can be characterized as a free boson field, so that
the results in section~\ref{ysec.3} may be utilized.  The general approach,
however, is applicable to any specific measurement system.

Consider the problem of estimating a parameter $U$ with Gaussian density
$p_G(u)$ of zero mean and variance $\sigma ^2$ via a single mode
optical field of average energy $S$.  While the optimization of (\ref{y7})
yields TCS as the solution, two choices have already been fixed in
advance: the parameter $u$ is to be modulated into the mean $\alpha _1$ of
the state, and homodyning or measurement of $\alpha _1$ is to be performed.
If one relaxes these two conditions in accordance with the general
quantum communication approach of section~\ref{ysec.3.1}, one may pick a state
$\rho(u)$ subject to
\beq
\int \mbox{d}up_G(u) tr\rho(u) a^{\dagger} a \leq S \label{y51}
\eeq
and a general measurement represented by the POM $\hat{O}(y)$, so that the
mean-square error $\ovr{\epsilon^2}$ between $y$ and $u$ is minimized.
It is not clear at all that the combination of linear modulation, 
TCS and homodyne is the
optimum solution or yields a near optimum performance to this problem.
As the following development shows, a lower bound for the
root-mean-square error $\delta u \equiv (\ovr{\epsilon^2})^{1/2}$
under (\ref{y51}) can be derived
\beq
\delta u \geq \sigma \frac{S^S}{(S+1)^{S+1}} \sim \frac{\sigma}{eS}
\; , \;\; S \gg 1 , \label{y52}
\eeq
which is very close to the TCS linear modulation performance,
\beq
\delta u^{TCS} = \frac{\sigma}{2S+1} \sim \frac{\sigma}{2S} \; , \;\;
S \gg 1 . \label{y53}
\eeq
Partly because it is not even clear whether the lower bound (\ref{y52}) can
indeed be achieved, one would ordinarily be quite satisfied with the
difference between 1/2 and $e^{-1}$ and stop looking for another system
unless the TCS system is not practical for whatever reason.  One may
say the linear TCS system is essentially optimum.  The corresponding
coherent state performance is $\delta u' \sim \sigma /\sqrt{S}$.

For a uniformly distributed phase parameter $\phi \in [-\pi ,
\pi)$, the corresponding lower bound for the root-mean-square error is
\beq
\delta \phi = \lambda \frac{S^S}{(S+1)^{S+1}} \sim \frac{\lambda}{eS} ,
\; \;  S \gg 1 , \label{y54}
\eeq
where $\lambda \sim 1.35$.  This single-mode $1/S$ behavior, improved
over the $1/\sqrt{S}$ dependence for coherent state, has been obtained
previously for two different concrete systems utilizing
TCS [60,61] and number states [62].  This should not be
surprising given the closeness between the number state and TCS
capacities,~(\ref{y31})-(\ref{y32}).

In the case of a narrowband field with $m = D/2 = WT$ modes but the
same total energy or photon number $S$, the lower bound for the
measurement of a Gaussian $U$ is
\begin{eqnarray}
\delta u &  \geq & \sigma \frac{S^S m^m}{(S+m)^{S+m}} \label{y55}\\
& \rightarrow & (1 + \frac{m}{S})^{-S}e^{-S} \; , \; \; m \gg 1 .
\label{y56}
\end{eqnarray}
For the uniform phase $\phi$, $\delta \phi$ is given by (\ref{y55}) with
$\sigma$ replaced by $\lambda$ as in (\ref{y52}) and (\ref{y54}).  Note
that (\ref{y56}) goes to zero as $D \rightarrow \infty$.  This is because
infinite capacity is obtained when the narrowband expression is
extrapolated to infinite bandwidth, which is not physically meaningful
as the $hf$ dependence becomes important [21,37]. The capacity is finite
when such dependency is taken into account [28,37-38].

The result (\ref{y55}) or (\ref{y56}) is rather remarkable. For the same total
energy $S$ spread over a large number of modes, the performance can be
improved from $1/S$ to $e^{-S}$ assuming the bound can be approached.
As a communication limit for which one can control the modulation,
this is certainly the case.  Intuitively, it can be traced to the
relative importance of $D$ or $W$ over $P$ or $S$ as discussed in
sections~\ref{ysec.2.1} and~\ref{ysec.3.2}. As the quantum state
affects only the SNR and
not $D$, one may expect such behavior for coherent states also, which
is indeed born out to be the case as developed in section~\ref{ysec.4.3}.
Even in the measurement situation, the improvement of a fixed energy spread
over many modes is a real one, to be demonstrated for a concrete
frequency modulation scheme also developed in section~\ref{ysec.4.3}.

(An aside between multimode and single mode results.  It is sometimes
said that two-mode squeezing is basically different from single mode
squeezing because two modes are needed.  However, by a simple modal
transformation equivalent to removing cross terms in the multimode
hamiltonian, two-mode squeezing can be reduced to single mode
squeezing, that is, in the multimode situation one picks the right
mode that yields squeezing. The general theory is given in [63,64],
which indeed was what led to the prediction of squeezing in degenerate
four-wave
mixing [65].)

\subsection{Classical rate-distortion limit}\label{ysec.4.2}

The rate-distortion function $R(d)$ of a random variable $U$ was
introduced by Shannon [11], with by now a very extensive
literature.  Here we consider just continuous $U$ with density
function $p(u)$ although discrete $U$ works the same.  For a
distortion measure $d(u,v)$ between $u$ and $v$, such as $|u-v|^2$ or
$|u-v|$, the average distortion is
\beq
E[d(U, V)] \equiv \int d(u,v)p(u)p(v|u)dudv . \label{y57}
\eeq
The rate distortion function $R(d)$ of $u$ is defined to be the
minimum mutual information
\beq
R(d) \equiv \min_{E[d(U,V)] \leq d} I(U; V) \label{y58}
\eeq
over all possible choices of $p(v|u)$ subject to the constraint that
$E[d(U, V)]$ is less than or equal to a given level $d \geq 0$.  One
may think of $V$ as a data-compressed version of $U$ --- $V$ represents $U$
with an average distortion $d$, thus it takes less bits to represent $V$
than $U$ for $d > 0$.  Shannon's source coding theorem with a fidelity
criterion and its converse [7,9,12] state that a source
variable $U$ can be asymptotically represented with an average
distortion $d$ if and only if at least $R(d)$ bits per source symbol
is provided.  Similar to channel coding, long sequence encoding and
decoding that ensures statistical regularity are in general required to
achieve such minimum in the asymptotic limit.  Nevertheless, roughly
speaking $R(d)$ is the minimum number of information bits per symbol
required to represent a source with an average distortion $d$ per
symbol.

The channel capacity $C$ may be written as a function $C(\beta )$
where $\beta$ denote the resource parameters available, including
power and bandwidth, as well as other characteristics of the channel
such as noise power.  Referring to Fig.~\ref{yufig1}, the question
arises on the
minimum distortion $d$ one can obtain for transmitting a source
variable $U$ over a channel with capacity $C(\beta )$.  The answer
is provided by Shannon's joint source-channel coding
theorem [7,9,12] in the so-called rate distortion limit or
rate distortion bound.  Recall that roughly speaking, the channel
coding theorem says that $C(\beta )$ is the maximum number of
information bits one can transfer error-free over a channel with
parameters $\beta$.  By combining the source and channel coding
theorem, one has $C(\beta ) \geq R(d)$ so that, since $R$ is a
monotone decreasing function of $d$,
\beq
d \geq R^{-1} C(\beta ) . \label{y59}
\eeq
Intuitively, this works as a lower bound as a consequence of the converse to the
coding theorem because otherwise one can transmit more than the
capacity rate or compress smaller than the source rate.  The positive
coding theorem assures that the bound may be approached arbitrarily
closely in a communications situation.

Even though the rate distortion bound (\ref{y59}) is generally achieved only
with source and channel coding, it is occasionally achieved without
any coding or nonlinear modulation.  Consider the transmission of a
zero-mean Gaussian $U$ of variance $\sigma ^2$ under with the
mean-square error criterion, $d(u,v) = |u-v|^2$, over an additive
Gaussian noise channel
\beq
X^{(out)} = X^{(in)} + n \label{y60}
\eeq
with noise variance $N$.  Then [7,9,11-13] for U
\begin{eqnarray}
R_u(d) & = & \frac{1}{2} \log (\sigma ^2/d), \qquad 0 \leq d \leq
\sigma^2  \label{y61}\\
       & = & 0, \qquad\qquad\qquad d \geq \sigma^2 \nonumber
\end{eqnarray}
and
\beq
C(S) = \frac{1}{2} \log(1 + \frac{S}{N}) \label{y62}
\eeq
under $E[ X^2] \leq S$.  The rate distortion limit (\ref{y59}) becomes
\beq
d \geq \sigma^2 (1+\frac{S}{N})^{-1} . \label{y63}
\eeq
If one sends $U$ as $X$ over the channel as in (\ref{y23}) so that $S = A^2
\sigma^2$, and use the estimate (\ref{y24}), the resulting mean-square error
(\ref{y25}) is exactly the lower limit (\ref{y63}).  This shows that for this
problem of transmitting a Gaussian parameter matched, in per use or
per symbol to a Gaussian noise channel, even in the full generality of
Fig.~\ref{yufig1} there is {\em nothing} that can do better than linear
modulation-demodulation!  Indeed, no way other than
through $R(d)$ has ever been successfully employed to show the
optimality of linear modulation in this problem.

The rate distortion function of the uniform phase variable $\phi$ is
difficult to evaluate exactly. However, the Shannon upper and lower
bounds [7,11] on $R(d)$ for $\phi$ differs only by about
0.3 bit per symbol.  Thus the upper bound
\beq
R_{\phi}(d) \leq \log \frac{1.35}{\sqrt{d}} \label{y64}
\eeq
would be used for $R_{\phi}(d)$.

There are two complications in the application of the rate disortion
bound to measurement problems. The first arises from the fact that in
a measurement one has little or no room for source coding as the
parameter $U$ is usually out of one's control before modulating
onto the physical channel input variable $X^{(in)}$. Thus, while (\ref{y59})
remains a limit, in general there may be no way to approach it. One may
try to replace $R(d)$ of (\ref{y61}) by some realistic $R(d)$ obtained with
whatever one can do to $U$, but contrary to what is stated in
ref~[56], it is not clear how such $R(d)$ may be
evaluated.  On the other hand, from experience the $R(d)$ for a
Gaussian parameter with different encoding criteria vary litle,
and so the exact form of $R(d)$ is not expected to make any major
difference in the final result (\ref{y59}).

The second problem is, I believe, more serious and closer to the heart
of the matter. It arises because no channel coding may be employed in
a typical measurement situation.  One can similarly try to replace
$C(\beta )$ by a mutual information $J(\beta )$ incorporating the
realistic limitations and freedom, which again seems hard to do.
This is an essential connection, however, because the modulation of
$U$ into $X$ represents how one physically couples $U$
into the measurement medium in the measurement system. It makes a
difference whether the optical field couples to $U$ via an
interferometer configuration or a source impressing configuration,
and e.g., whether the frequency or the amplitude is modulated by
$U$.  In any case, if such a meaningful $J(\beta )$ can be
obtained, then a measurement rate distortion limit can be obtained
from $J(\beta ) \geq R(d)$ in the form similar to (\ref{y59})
\beq
d \geq R^{-1} J(\beta ) . \label{y65}
\eeq

\subsection{Ultimate quantum measurement system limit}\label{ysec.4.3}

By combining the above rate distortion theory with classical
capacities replaced by quantum capacities, one obtains quantum rate
distortion limits for a general quantum system of Fig.~\ref{yufig3},
taking into
account all the freedom of classical modulation-demodulation and
quantum measurement as well as state selections.  Note that the
uncertainty principles are far from sufficient to determine such
ultimate limits.  In the original form they are merely restrictions on
quantum states, and even in their extended form [20] they
do not account for the many freedom represented in Fig.~\ref{yufig3}.
A few more remarks on this may be found in [10] and [59].

>From (\ref{y61}) and the single mode version of (\ref{y32}), one finds (\ref{y52})
using the optimum number state capacity.  With $C_{TCS}$ of (\ref{y32}), one
finds the same $\delta u^{TCS}$ as (\ref{y53}) for linear modulation
without coding!  This is {\em exactly} the situation around (\ref{y25}) and
(\ref{y63}) pointed out above.  For $C_{het}$ of (\ref{y30}), one obtains
\beq
\delta u^{CS} = \frac{\sigma}{S+1} , \label{y66}
\eeq
which may be compared to
\beq
\delta u' \sim \frac{\sigma}{\sqrt{S}} \label{y67}
\eeq
obtained in coherent state systems without coding or nonlinear
modulation.  The reason why coding or nonlinear modulation is
necessary in the coherent state case is that bandwidth expansion, two
quadratures in a coherent state versus the single real parameter to be
estimated, has to be utilized.  Thus, apart from a gain on $\delta u$
by a factor of
2, {\em the use of TCS for measurement is essentially the same as
coding on a coherent state system} as far as the performance goes, a
rather unexpected result.

For the uniform phase parameter, one finds (\ref{y54}) from (\ref{y31})
and (\ref{y64}), and similarly
\begin{eqnarray}
\delta \phi ^{TCS} & \sim & \frac{1}{2S}, \; \; S \gg 1 \label{y68}\\
\delta \phi^{CS} & \sim & \frac{1}{S} ,\; \; S \gg 1 . \label{y69}
\end{eqnarray}
Again, the $1/S$ behavior can be obtained without coding on TCS or
number state systems, while coherent state systems without coding
yields
\beq
\delta \phi' \sim 1/\sqrt{S} . \label{y70}
\eeq

In the multimode narrowband situation, one similarly obtains (\ref{y55}) for
the optimum case, with $m = D/2 = WT$, and
\begin{eqnarray}
\delta u^{TCS} & = & \sigma (1 + \frac{2S}{m})^{-m} \label{y71}\\
\delta u^{CS} &  = & \sigma (1 + \frac{S}{m})^{-m} . \label{y72}
\end{eqnarray}
Similarly results for $\phi$ can be written down with $\sigma$
replaced by $\lambda$ as in (\ref{y52}) and (\ref{y54}).
>From (\ref{y71}) and (\ref{y72}),
\beq
\delta u ^{TCS} \rightarrow \sigma e^{-2S}, \hspace*{.2in} \delta u^{CS}
\rightarrow \sigma e^{-S}, \hspace*{.3in} D \rightarrow \infty \label{y73}
\eeq
an exponential decrease in $S$ versus $1/S$ in the single mode case.
Even though the narrowband assumption is violated in $D \rightarrow
\infty$, the exponential improvement is real from (\ref{y55}) or
(\ref{y71})- (\ref{y72}) as $D$ can be very large at optical frequencies.

To show that multimode system is indeed better for measurement for the
same total energy, consider the following pulse frequency modulation
scheme
\beq
X^{(in)}(t,\phi) = \sqrt{\frac{2S}{T}} \sin (w_0 + \beta \phi )t,
\qquad 0\leq t \leq T , \label{y74}
\eeq
where $\beta$ is a known fixed constant and $S$ the total energy in the
signal. Classically, it is known [8] that in the
presence of additive white Gaussian noise, the use of (\ref{y74}) and
corresponding nonlinear demodulation lead to a decrease of
root-mean-square error $\delta \phi$ by a factor $\sim \frac{1}{m}$
compared to the linear modulation case, when a threshold constraint
involving $S$, $T$, $\beta$ and the noise variance $N_0$ is satisfied
which occurs for sufficiently large $D$ or $S$.  If (\ref{y74}) is used in
either a coherent state-heterodyne or TCS-homodyne systems, one would obtain
\beq
\delta \phi^{TCS} \sim \frac{1}{mS}, \hspace*{.2in} \delta \phi^{CS} \sim
\frac{1}{m\sqrt{S}} \label{y75}
\eeq
compared to (\ref{y68}) and (\ref{y70}).  While showing the importance of
bandwidth, the net gain $1/m$ is in itself already significant as $D$
is large.

\section{Position monitoring with contractive states}\label{ysec.5}

As a final application of squeezed states, we discuss the problem of
repeatedly measuring the position of a free mass for which the state
after each measurement is important as it determines the state at the
next measurement instant.  This feature makes the problem, relevant to
gravitational-wave interferometers treated elsewhere in this book,
quite different from the other ones we have discussed so far in this
chapter, for which all the information can be extracted from the
system by one measurement.  There is still considerable confusion in
the literature on the validity of the so-called ``standard quantum
limit'' (SQL) on how small the position fluctuation $\lang \Delta
\hat{X}^2(t)\rang$ can be obtained in a sequence of position
measurements, although the issues in principle have been cleared up
entirely over ten years ago.  Perhaps this is partly because some of
the following clarification never appeared in print.

The SQL states that [66,67] if a position measurement is made at
$t=0$, the fluctuation at $t>0$ is at least
\beq
\lang \Delta \hat{X}^2(t)\rang_{SQL} = \hbar t/m , \label{y76}
\eeq
where $m$ is the mass of a fermion.  The derivation of (\ref{y76}), however,
was incorrectly taken to be universally valid as a consequence of the
Uncertainty Principle, and it was concluded that the free mass
position is not a ``QND observable'' --- namely, that the disturbance
to the system from the first position measurement demolishes the
possibility of an accurate second measurement after an interval of
free evolution.  To delineate how a position monitoring scheme works,
consider the monitoring of weak classical forces $f_1(t), f_2(t)$
coupled linearly to a free mass with position $\hat{X}$ and momentum $\hat{P}$,
\beq
H_I = f_1(t)\hat{X}+f_2(t)\hat{P} . \label{y77}
\eeq
In the Heisenberg picture,
\beq
\hat{X}(t) = \hat{X}(0)+ \hat{P}(0)t/m + \int^t_0 f_2 (t') dt' -
\int^t_0 dt' \int^{t'}_0 d\tau f_1(\tau )/m , \label{y78}
\eeq
\beq
\hat{P}(t) = \hat{P}(0) - \int^t_0 f_1(t')dt' . \label{y79}
\eeq
Typically, $f_2 = 0$ and $\hat{X}$ is more readily measureable than
$\hat{P}$ in practice.  From (\ref{y78}), information on $f_1(t)$ can be
obtained by measurements on $\hat{X}(t)$ at different times.

If a position measurement at $t=0$ is made in the sense of Pauli's
first-kind measurement [68], the position eigenstates $|X\rang$ is
to be used to compute the measurement probability $p(X)$ and the state
of the mass after the measurement with a reading $X'$ is $|X'\rang$.
Thus, first-kind measurement of a selfadjoint operator is one for
which the Von Neumann projection postulate applies.  From (\ref{y79}), the
position fluctuation $\lang \Delta \hat{X}^2(t)\rang$ is often
concluded to be infinite, because the ``back-action'' causes $\lang
\Delta ^2 \hat{P} (0) \rang = \infty$ with $\lang \Delta \hat{X}^2 (0)
\rang = 0$ from the Uncertainty Principle.  Since $\lang
\hat{P}^2(0)\rang = \lang \Delta ^2 \hat{P}(0) \rang + \lang
\hat{P}(0)\rang ^2$, an infinite average energy is obtained for the
mass in a position eigenstate $|X\rang$, thus one can actually only
make ``approximate'' position measurements which are generally
described by POM as far as the measurement statistics goes, with
$\lang \Delta \hat{X}^2(0)\rang > 0$.  In any event, it was concluded
that whatever position measurement is used the Uncertainty Principle
implies the SQL (\ref{y76}).

In [69], it was pointed out that this conclusion is not valid from
(\ref{y78}) when $\lang \Delta \hat{X}(0)\Delta \hat{P} (0)+\Delta \hat{P}
(0)\Delta \hat{X}(0) \rang$ is negative.  It was also pointed out that
$\lang \Delta \hat{X}^2 (t) \rang$ can be arranged to be as small as
desired at any $t > 0$ if the state after measurement is left in a
``contractive state'' $|\mna \omega \rang$, which is a TCS $| \mna
\rang$ with the frequency $\omega$ put back explicitly and the
parameters $\mu, \nu, \omega$ chosen appropriately so that the
``generalized minimum uncertainty wave packet'' $\lang X|\mna \omega
\rang$ contracts rather than spreads in $t$ up to a desired
measurement time.  It was observed that measurements of the second
kind, in particular a class of measurements formally described by
Gordon and Lonisell, may be used to beat the SQL. Specifically, the
measurement described by [68,69]
\beq
|\mna \omega \rang \lang \mu ' \nu ' \alpha \omega | \label{y80}
\eeq
would work, where $|\mu ' \nu ' \alpha \omega \rang$ is used to compute
the measurement probability with reading $\alpha = \alpha_1 +
i\alpha_2$,
\beq
\alpha_1 = x (m\omega /2\hbar )^{1/2} , \hspace*{.2in} \alpha_2 = p/(2
\hbar m \omega)^{\frac{1}{2}} , \label{y81}
\eeq
which may be considered a joint approximate measurement of $\hat{X}$
and $\hat{P}$ similar to TCS-heterodyne [22], and $|\mna \omega \rang$
is the state after measurement of reading $\alpha$ arranged to be a
contractive state for the next measurement.  The position measurement
would be sharp if $\lang \Delta \hat{X}^2\rang \sim |\mu ' -
\nu '|^2 \rightarrow 0$, while $\mu , \nu , \omega$ are chosen so that
the mass state has a sharply defined position at the next measurement
instant.

Two criticisms were made on the success of this approach to beat the
SQL.  First, it was pointed out that it was not clear a measurement
described as in (\ref{y80}) is realizable in principle.  A quantum
measurement realization can be described by the coupling of a
``proble'' to the system with commuting selfadjoint operators being
measured on the probe, and with all the quantities computed by the
usual rules of quantum mechanics (without the need for the projection
postulate as emphasized by Ozawa [16].)  While two realizations
were produced [71], they were criticized on the ground that the
probe-system interaction hamiltonians $H_I$ are time-dependent and so
are equivalent to ``state preparation.''  While these realizations are
actually quite different from the state preparations that were
discussed and are in fact full-fledged quantum measurement
realizations in accordance with standard quantum measurement theory,
the situation is resolved beyond dispute when a time-independent $H_I$
was found [72] for realizing (\ref{y80}).  More significantly,
Ozawa [16,73] has obtained a complete characterization of quantum
measurement including the state after measurement in the concept of a
completely positive operation measure mentioned in Section~\ref{ysec.3.1},
and he showed that any Gordon-Lonisell measurement representation, in
which a complete but not necessarily orthogonal set of states is used
to yield the measurement statistics and the state after measurement
depends only the measured value, is indeed realizable.

To discuss the second criticism, one needs to examine more closely how
the measurement scheme based on (\ref{y80}) actually works.  Let $\alpha '$
be the reading at $t = 0$ so that the state at $t = 0+$ 
is $|\mna' \omega \rang$.  After another time $t$, the free mass is in
state $|\mu_t \nu_t \alpha'_t \omega \rang$ with $|\mu_t - \nu_t |
\rightarrow 0$.  From (\ref{y78})-(\ref{y79}) with $f_2 = 0$, the value
$\alpha '_t$ is given by
\begin{eqnarray}
\alpha '_{t1} & = & \alpha '_1 + \alpha '_2 t/m - \int^t_0 dt'
\int^{t'}_0 d\tau f_1 (\tau) /m \label{y82}\\
\alpha'_{t2} & = & \alpha'_2 -  \int^t_0 f_1 (\tau)d\tau . \label{y83}
\end{eqnarray}
Equations (\ref{y82}) and (\ref{y83}) provide the average of the reading
$\alpha^{''}$ at $t$, which can be represented by
\begin{eqnarray}
\alpha^{''}_1 & = &  \alpha'_1 + \alpha'_2 t/m - \int^t_0 dt'
\int^{t'}_0 d\tau f_1(\tau )/m + n_1 , \label{y84}\\
\alpha^{''}_2 & = & \alpha'_{t2} + n_2 , \label{y85}
\end{eqnarray}
where the fluctuation of $n_1$ is vanishingly small from $|\mu_t -
\nu_t| \rightarrow 0$ while the noise $n_2$ is big.  From (\ref{y84}), one
may use the $\alpha^{''}_1$ reading to estimate $f_1$ after it is
subtracted from the value of $\alpha'_1 + \alpha'_2 t/m$ known at time
$t$.  The reading $\alpha^{''}_2$ is also taken so that it could be used
for the subtraction at the next measurement, although it is not used
for estimating $f_1$ as it is noisy and helps little.  It is clear
that this scheme beats the SQL to any arbitrary level in a sequence of
measurements.

In [74], a ``predictive sense'' of the SQL was proposed to
suggest that the SQL was not beaten in that sense.  This predictive
sense can be described by the stipulation that prior to any
measurement, $\alpha'_1$ and $\alpha'_2$ in (\ref{y84}) and (\ref{y85})
are unknown and
random, thus $\alpha^{''}_1$ is also more random than $n_1$ and indeed
obeys the SQL.  But since we know we {\em will} have the reading value
$\alpha'$ available at $t$ which would be subtracted from (\ref{y84}), the
reading $\alpha^{''}$ at $t$, we {\em can indeed predict} we will get
$\lang \Delta \hat{X}^2(0)\rang$, $\lang \Delta \hat{X}^2 (t)\rang$, and so on,
arbitrarily small.  Thus, the SQL is beaten by (\ref{y80}) in the predictive
sense.  Further eludication of this point and discussion
on the working of this scheme (\ref{y80})-(\ref{y85}) was provided in [75].

Actually, this issue would not even arise if the measurement
\beq
|\mu \nu 0 \omega \rang \lang \mu ' \nu ' \alpha \omega | \label{y86}
\eeq
is employed instead of (\ref{y80}), for which the state after measurement
always has $\lang \hat{X} \rang = \lang \hat{P} \rang = 0$.  This
measurement is a special degenerate case of Gordon-Lonisell
measurement, and thus realizable by Ozawa's theorem.  Indeed, an
explicit hamiltonian realization can be developed for (\ref{y86}) [76,77].

Since the positions of a free mass can be repeatedly measured
accurately, it is not appropriate to say that $\hat{X}$ is not a QND
observable.  The term QND measurement is often used just to refer to a
first-kind measurement, which is an acceptable terminology.  What has
never been demonstrated is that there is, in principle, any observable
which is not a QND observable in the generic sense.  In fact, it
should be clear from the development in this section, and it can
indeed be readily shown in principle, that any observable can be
repeatedly measured arbitrarily accurately in the absence of
particular constraints.  The key point is that, as in (\ref{y80}), the state
used to compute the measurement probability and the state after
measurement need not be the same.

\end{document}